\documentclass[aps,amsmath,onecolumn,amssymb,nofootinbib,floatfix,showpacs]{revtex4}
\usepackage{graphicx}
\usepackage{amsmath}
\usepackage{amssymb}
\usepackage{latexsym}
\usepackage{dcolumn}
\usepackage{bm}
\usepackage{hyperref} 
\usepackage{url}
\tighten

\begin{document}
\title{Soliton solutions of the improved quark mass density-dependent model at finite temperature }

\author{Hong Mao$^1$}
\email {maohong@hp.ccast.ac.cn}
\author{Ru-Keng Su$^{1,2}$}
\author{Wei-Qin Zhao$^{1,3}$}
\address{1. CCAST(World Laboratory), P.O. Box 8730, Beijing 100080,
China  \\
2. Department of Physics, Fudan University,
Shanghai 200433, China\\
3.Institute of High Energy Physics, Chinese Academy of Sciences,
Beijing 100049, China}


\begin{abstract}
The improved quark mass density-dependent model (IQMDD) based on
soliton bag model is studied at finite temperature. Appling the
finite temperature field theory, the effective potential of the
IQMDD model and the bag constant $B(T)$ have been calculated at
different temperatures. It is shown that there is a critical
temperature $T_{C}\simeq 110 \mathrm{MeV}$. We also calculate the
soliton solutions of the IQMDD model at finite tmperature. It turns
out that when $T<T_{C}$, there is a bag constant $B(T)$ and the
soliton solutions are stable. However, when $T>T_{C}$ the bag
constant $B(T)=0$ and there is no soliton solution, therefore, the
confinement of quarks are removed quickly.
\end{abstract}


\maketitle

\section{Introduction}
It is widely believed that the fundamental theory of strong
interaction is quantum chromodynamics (QCD), which in principle can
be applied to describe most of nuclear physics. However, due to the
property of confinement and asymptotic freedom, QCD can not yet be
used to study low-energy nuclear physics. The challenge to nuclear
physicists is to find models which can bridge the gap between the
fundamental theory and our wealth of knowledge about low energy
phenomenology. Some of these models have been proved to be
successful in reproducing different properties of hadrons, nuclear
matter and quark matter
\cite{Serot:1997xg,Tomas:1984,Farhi:1984qu,Madsen:1989pg,Greiner:1987tg,Fowler:1981rp}.
The quark mass density-dependent model (QMDD) is one of such
candidates.

The QMDD model was first suggested by Fowler, Raha and Weiner
\cite{Fowler:1981rp}, and then employed by many authors to study the
stability and physical properties of strange quark matter
\cite{Chakrabarty:1991ui,Benvenuto:1989kc,Peng:2000ff,Wang:2000dc}.
This model provides an alternative phenomenological description of
quark confinement because in this model, the mass of quarks is given
by
\begin{eqnarray}
m_{u,d} &=& \frac{B}{n_{q}}, \label{mass_ud}\\
 m_s  &=& m_{s0} +\frac{B}{n_{q}},\label{mass_s}
\end{eqnarray}
where $B$ is the vacuum energy density and $n_q=n_u+n_d+n_{s0}$.
$n_u, n_d, n_{s0}$ represent the number density of the $u, d$ and
$s$ quarks respectively. We see from Eqs.(\ref{mass_ud}),
(\ref{mass_s}), if the distance between quarks goes to infinity, the
volume of the system becomes infinite, $n_q$ approaches to zero and
$m_q$ ($q=u,d,s$) approaches to infinite. This is just the
confinement condition as that of MIT bag model \cite{Tomas:1984}.
However, QMDD model has its advantage since the boundary condition
put by hand in MIT model to confine quarks is abandoned. This
correspondence was confirmed by Benvenuto and Lugones
\cite{Benvenuto:1989kc}. They proved that the properties of strange
matter in the QMDD model are nearly the same as those obtained in
the MIT bag model.

Although the QMDD model can provide a dynamical description of
confinement and explain the stability and many other dynamical
properties of strange quark matter (SQM) at zero temperature, there
are many difficulties when we extend this model to finite
temperature. They are: 1) it cannot mimic the correct "temperature
$T$ vs. density $\rho$" deconfinement phase diagram of QCD because
the quark masses are divergent and then $T$ tends to infinite when
$n_q \to 0$ \cite{Zhang:2002fs}; 2) Although the basic improvement
of QMDD model is the quark masses depending on density and the quark
confinement mechanism mimics, it is still an ideal quark gas model
with no quark-quark interaction included. However, as was shown in
recent RHIC experiments, the quark-quark interaction cannot be
neglected in Quark-Gluon plasma. In order to overcome the first
difficulty, a quark mass density- and temperature dependent (QMDTD)
model
\cite{Zhang:2002fs,Zhang:2002qq,Zhang:2003ka,Zhang:2003fn,Zhang:2001ih}
is proposed and it was argued that $B$ is a function of temperature
as that of Friedberg-Lee (FL) model
\cite{Friedberg:1976eg,Birse:1991cx,Wilets:1990di}. The equation of
state of QMDTD model are used to study the properties of strange
quark star in Refs.\cite{Gupta:2002hj}\cite{Shen:2005vh} and the
results are successful. To the second difficulty, as a first step,
Ref.\cite{Wu:2005ty} has introduced a coupling between quark and
non-linear scalar field to improve the QMDD model and have
investigated the properties of nucleon successfully. Instead of the
correspondence of MIT bag to QMDD model, a FL soliton bag is formed
by quark and scalar field coupling for improved QMDD (IQMDD) model
and the quark deconfinement phase transition can take place.

This paper evolves from an attempt to extend the IQMDD model
\cite{Wu:2005ty} to finite temperature. Since the spontaneously
broken symmetry of the scalar field will be restored at finite
temperature, the non-topological soliton tends to disappear when
temperature $T$ increases to $T_{C}$. If the temperature increases
further, the solution becomes a damping oscillation and such a
solution can not be taken as "soliton " solution, so the quarks are
to be deconfined.

By using the finite temperature field theory, we will calculate the
effective potential under one-loop approximation and find $T_{C}
\approx 110 \mathrm{MeV}$. Studying the temperature dependence of
soliton solutions is our first motivation. Our second motivation is
to find the function $B(T)$. In QMDTD model, $B(T)$ is an input. As
an ansatz, two formulae
\begin{equation} \label{e3}
B_1 (T) = B_0 [1 - A_1(\frac{T}{{T_C }}) + A_2(\frac{T}{{T_C }})^2
],\quad 0 \le T \le T_C
\end{equation}
with $A_1,A_2$ adjustable parameters, and
\begin{equation} \label{e4}
B_2 (T) = B_0 [1 - (\frac{T}{{T_C }})^2 ], \quad 0 \le T \le T_C
\end{equation}
were introduced in Ref.\cite{Zhang:2001ih} and
Ref.\cite{Zhang:2002fs} respectively. However, for FL model or for
the IQMDD model, $B(T)$ can be calculated because the vacuum energy
density $B$ equals the difference of the value of the effective
potential between the vacua inside and outside the soliton bag, and
the vacua of the effective potential depend on temperature.

The organization of this paper is as follows. In the following
section we review the IQMDD model and its numerical solutions for
our selected parameters. In Sec.III, we give detailed calculation of
the effective potential of the IQMDD model and the bag constant as
functions of temperature. The soltion solutions of the IQMDD model
at different temperatures are presented in Sec.IV, while in the last
section we present our summary and discussion.

\section{The IQMDD model}

Since the details of IQMDD model can be found in
Ref.\cite{Wu:2005ty}, hereafter we only write down the essential
formulae which are necessary for our further discussion. The
effective Lagrangian density of the IQMDD model is \cite{Wu:2005ty}
\begin{eqnarray}\label{lagrangian}
\mathcal{L}=\overline{\psi}(i\eth -g \sigma
-m_{q})\psi+\frac{1}{2}\partial_{\mu}\sigma \partial^{\mu}\sigma
-U(\sigma),
\end{eqnarray}
which describes the interaction of the spin-$\frac{1}{2}$ quark
fields $\psi$ and the phenomenological scalar field $\sigma$ with
the coupling constant $g$. $m_q$ is the mass of u(d) quark, which is
given by Eq.(\ref{mass_ud}). In this paper we do not discuss the
strange quark. The potential for the $\sigma$ field is chosen as
\begin{eqnarray}
U(\sigma)=\frac{a}{2!}\sigma^2+\frac{b}{3!}\sigma^3+\frac{c}{4!}\sigma^4+B,
\end{eqnarray}
\begin{equation} \label{e7}
b^2 > 3ac
\end{equation}
The condition (\ref{e7}) ensures that the absolute minimum of
$U(\sigma)$ is at $\sigma = \sigma_{v} \ne 0$.

$U(\sigma)$ has two minima: one is the absolute minimum at a large
value of the $\sigma$ field
\begin{eqnarray}
\sigma_v=\frac{3|b|}{2c}\left[1+\left[1-\frac{8ac}{3b^2}\right]^{\frac
1 2}\right],
\end{eqnarray}
with $U(\sigma_v)=0$, another is at $\sigma_0=0$. The former
corresponds to the physical or nonperturbative vacuum, and
$\sigma=\sigma_v$ representing some condensates. The other
represents a metastable vacuum where the condensates vanishes, with
an energy density $B$ relative to the physical vacuum. The "bag
constant" $B$ can be expressed as
\begin{eqnarray}
-B=\frac{a}{2!}\sigma^2_v+\frac{b}{3!}\sigma^3_v+\frac{c}{4!}\sigma^4_v.
\end{eqnarray}

From Eq.(\ref{lagrangian}), the classical Euler-Lagrangian equations
can be obtained as
\begin{eqnarray}\label{euler-eq1}
(i\eth-g\sigma-m_q)\psi &=& 0, \\ \square
\sigma+\frac{dU}{d\sigma}+g\overline{\psi}\psi &=& 0.
\label{euler-eq2}
\end{eqnarray}
In the mean-field approximation (MFA), the scalar field $\sigma$ is
taken as time-independent classical c-number field, and we only
consider a fixed occupation number of valence quarks (3 quarks for
nucleons, and quark-antiquark pair for mesons). Quantum fluctuation
of the bosons and effects of the quark Dirac sea are thus to be
neglected. In the following , we will discuss the ground state
solution of the system.

In the spherical case, the $\sigma$ is spherically symmetric, and
valence quarks are in the lowest s-wave level. Then the scalar field
$\sigma$ and the Dirac equation functions can be written as
\begin{eqnarray}
\sigma(\mathbf{r},t) &=& \sigma(r), \\
\psi(\mathbf{r},t) &=& e^{-i\epsilon t}\sum_{i}\varphi_i,
\end{eqnarray}
where the quark Dirac spinors have the form
\begin{eqnarray} \label{spinor}
\varphi=\left(\begin{array}{c} u(r) \\
i \vec{\sigma} \cdot\mathbf{\hat{ r}}v(r)
\end{array}\right)\chi.
\end{eqnarray}
By using Eqs.(\ref{euler-eq1})-(\ref{spinor}), we obtain
\begin{eqnarray}
\frac{du(r)}{dr}=-(\epsilon+m_q+g \sigma(r))v(r), \label{equation1}\\
\frac{dv(r)}{dr}=-\frac{2}{r}v(r)+(\epsilon-m_q-g \sigma(r))u(r), \label{equation2} \\
\frac{d^2
\sigma(r)}{dr^2}+\frac{2}{r}\frac{d\sigma(r)}{dr}-\frac{dU}{d\sigma}=Ng(u^2(r)-v^2(r)).
\label{equation3}
\end{eqnarray}
The quark functions should satisfy the normalisation condition
\begin{eqnarray}
4\pi \int r^2 (u^2(r)+v^2(r))dr=1.
\end{eqnarray}
The number of quarks would be $N=3$ for baryons and $N=2$ for
mesons. In the following discussions, we only constrain in the case
$N=3$. These equations are subject to the boundary conditions which
follow from the requirement of finite energy:
\begin{eqnarray}
v(0)=0,     \frac{d\sigma(0)}{dr}=0, \nonumber\\
u(\infty)=0,          \sigma(\infty)=\sigma_v \nonumber.
\end{eqnarray}

If we consider $N$ quarks in the lowest mode with energy $\epsilon$,
the total energy of the system is
\begin{eqnarray}\label{energy}
E=N \epsilon+4\pi\int r^2 \left[ \frac{1}{2}
\left(\frac{d\sigma}{dr}\right)^2+U(\sigma) \right]dr.
\end{eqnarray}

This model has four adjustable parameter $g,a,b,c$ which can be
chosen to fit various baryon properties such as the proton change
radius $r_{cp}$, the proton magnetic moment $\mu_{p}$ and the ratio
of axial-vector to vector coupling $g_A / g_v$. In
Ref.\cite{Wu:2005ty}, a wide rage of parameters has been used to
calculate above quantities. Hereafter we take one set of parameters
$a=70.0 fm^{-2}$, $b=-2201.8 fm^{-1}$, $c=20000$ and $g=9.8$ to
study the temperature dependence of the soliton solution. It has
been proven in Ref.\cite{Wu:2005ty} that this set of parameters can
describe the properties of nucleon at zero temperature successfully.

\begin{figure}
\includegraphics[scale=0.9]{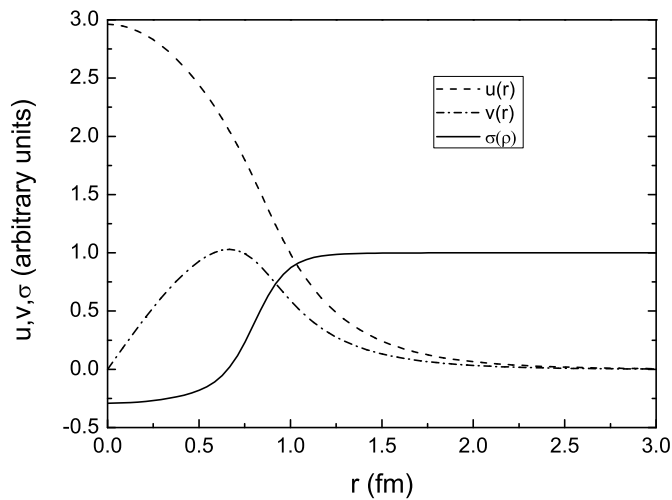}
\caption{\label{Fig:Fig1} The $\sigma$ and quark fields in arbitrary
unit as functions of r for the parameters taken as $a=70.0 fm^{-2}$,
$b=-2201.8 fm^{-1}$, $c=20000$ and $g=9.8$.}
\end{figure}

\begin{figure}
\includegraphics[scale=0.9]{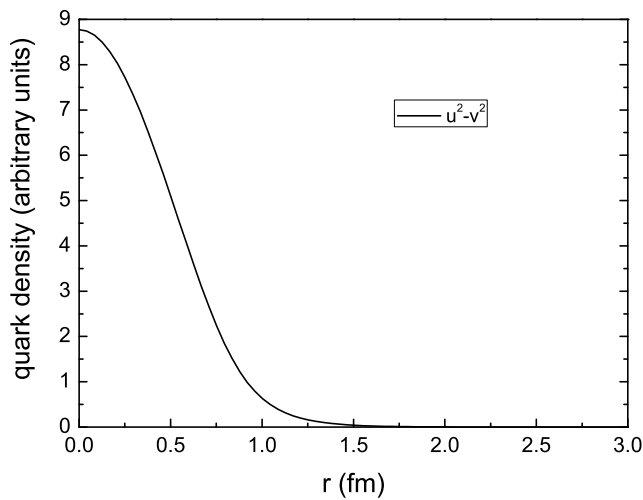}
\caption{\label{Fig:Fig2} The quark density $u^2(r)-v^2(r)$ in
arbitrary unit as functions of r for the parameters taken as $a=70.0
fm^{-2}$, $b=-2201.8 fm^{-1}$, $c=20000$ and $g=9.8$.}
\end{figure}

In Fig.\ref{Fig:Fig1} we plot the $\sigma$ and quark fields profiles
in arbitrary unit as functions of r for the above parameters. The
quark density $u^2(r)-v^2(r)$ distribution versus radius are plotted
in Fig.\ref{Fig:Fig2}.

\section{Effective potential at finite temperature}

The appropriate framework for studying phase transitions is finite
temperature field theory or thermal field theory. Within this
framework the finite temperature effective potential is an important
and useful theoretical tool. The idea of such techniques could trace
back to the 1970's when Kirzhnits and Linde first proposed that
symmetries broken at zero temperature could be restored at finite
temperature\cite{Kirzhnits:1972iw}\cite{Kirzhnits:1972ut}.
Subsequently Weinberg\cite{Weinberg:1974hy}, Dolan and Jackiw
\cite{Dolan:1973qd}as well as many others have adopted the effective
potential as the basic tool to study the symmetry-breaking phase
transitions. In this section we briefly outline the relevant results
of Dolan and Jackiw\cite{Dolan:1973qd}, since the IQMDD model is
very similar to the model studied by those authors.

The effective potential can be calculated to one-loop order by using
the methods of Dolan and Jackiw\cite{Dolan:1973qd}. But it is
pointed out in Ref\cite{Friedberg:1976eg, Lee:1981mf} that, as an
approximation, all $\sigma$ quantum loop diagrams may be ignored due
to the fact that $\sigma$ is only a phenomenological field
describing the long-range collective effects of QCD, its short-wave
components do not exist in reality. Therefore for the rest of this
discussion we shall ignore quantum corrections and concentrate on
those induced by finite temperature effects. Then the one-loop
contribution to the effective potential is of the form
\begin{eqnarray}\label{potential0}
V(\sigma;\beta)=U(\sigma)+V_B(\sigma;\beta)+V_F(\sigma;\beta),
\end{eqnarray}
where $U(\sigma)$ is the classical potential of the Lagrangian.
$V_B(\sigma;\beta)$ and $V_F(\sigma;\beta)$ are the finite
temperature contributions from boson and fermion one-loop diagrams
respectively\cite{Dolan:1973qd}\cite{Das:1997gg}. It should be
pointed out that, in this approximate method for the calculation of
the effective potential, the plane wave quark states have been used,
while, in fact, the quarks are confined in the finite-size solitonic
configuration. Therefore, this approximate method is not fully
self-consistent. For simplicity, we have substituted $\sigma(T)$ for
$\bar{\sigma}(T)$. These contribute the following terms in the
potential\cite{Dolan:1973qd}
\begin{eqnarray}\label{potential1}
V_B(\sigma;\beta)=\frac{1}{2\pi^2 \beta^4} \int^{\infty}_0 dx x^2
\mathrm{ln} \left( 1-e^{-(x^2+\beta^2 m_{\sigma}^2)} \right),
\end{eqnarray}
\begin{eqnarray}\label{potential2}
V_F(\sigma;\beta)=-12\sum_n \frac{1}{2\pi^2 \beta^4} \int^{\infty}_0
dx x^2 \mathrm{ln} \left( 1+e^{-(x^2+\beta^2 m_{qn}^2)} \right),
\end{eqnarray}
where the minus sign is the consequence of Fermi-Dirac statistics.
$m_{\sigma}$ and $m_q$ are the effective masses of the scalar field
$\sigma$ and the quark field, respectively:
\begin{eqnarray}
m_q &=& \frac{B(T)}{n_q}+g \sigma(T), \label{massq}\\
m^2_{\sigma}&=& a+b \sigma(T)+\frac{c}{2} \sigma^2(T). \label{masss}
\end{eqnarray}
We fix $m^2_{\sigma}$ by taking its value at the physical vacuum
state\cite{Gao:1992zd}. In this work, $B(T)$ is defined as the
difference between the vacua of the effective potential inside and
outside the solion bag. This means that for $T\leq T_{C}$, $B(T)$ is
the difference between the effective potential values at the
perturbative vacuum state and values at the physical vacuum state
\begin{eqnarray}\label{bag}
B(T)=V(\sigma_0;\beta)-V(\sigma_v;\beta).
\end{eqnarray}
For $T > T_{C}$, $B(T)=0$ due the fact that the vacua inside and
outside the soliton bag are equal. This will be analyzed in more
details in next section.

\begin{figure}
\includegraphics[scale=0.9]{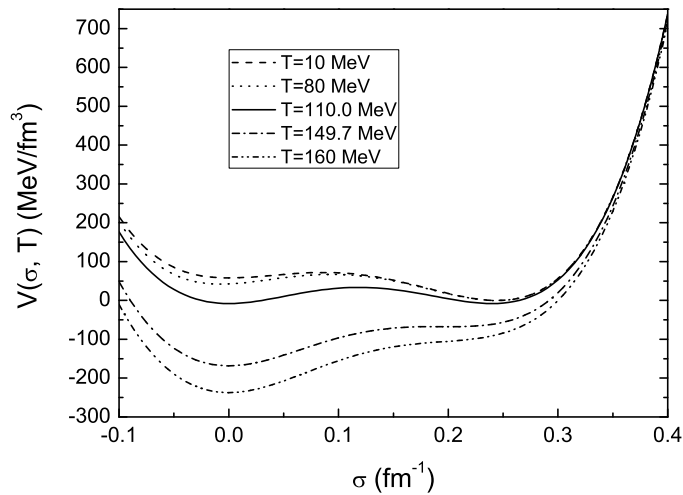}
\caption{\label{Fig:Fig3} The temperature-dependence of the one-loop
effective potential $V(\sigma;\beta)$. At low temperature, there are
two minima and one of which is going to disappear when the
temperature reaching a high temperature $T=149.7 \mathrm{MeV}$. The
critical temperatures is set at $T_{C}\simeq 110 \mathrm{MeV}$ when
two minima are euqal.}
\end{figure}

From Eqs.(\ref{potential0}), (\ref{massq}) and (\ref{masss}), we can
numerically solve the effective potential $V(\sigma;\beta)$ for
different temperatures. In Fig.\ref{Fig:Fig3}, we plot the one-loop
effective potential $V(\sigma;\beta)$ as a function of $\sigma$ at
$T=10 \mathrm{MeV}$, $T=80 \mathrm{MeV}$, $T=110.0 \mathrm{MeV}$,
$T=149.7\mathrm{MeV}$ and $T=160 \mathrm{MeV}$. It can be seen from
Fig.\ref{Fig:Fig3} that there exist two particular temperatures. One
is that the effective potential exhibits two degenerate minima at
$T_{C}\simeq 110 \mathrm{MeV}$ which is defined as critical
temperature, the other is that the second minimum of the potential
at $\sigma\simeq \sigma_v$ disappears at a higher temperature
$T\simeq 149.7 \mathrm{MeV}$.

For low temperatures the absolute minimum of $V(\sigma;\beta)$ lies
close to $\sigma_v$, and there is another minimum at $\sigma_0$. The
physical vacuum state at $\sigma_v$ is stable and correspondingly
quarks are in confinement. As the temperature increases the second
minimum of the potential at $\sigma_0$ decreases relative to the
first one. At the critical temperature $T_{C}\simeq 110
\mathrm{MeV}$, the potentials at the two minima are equal. The
physical vacuum becomes unstable. Since $B(T)=0$ when $T$ is lager
than $T_{C}$, there is no bag constant to provide the dynamical
mechanism to confine the quarks in the bag. Moreover, there is no
soliton-like solution in the model when the temperature is above
$T_C$ due to the effective potential and vacuum structure, there
only exists the damping oscillation solution and such a solution can
not produce a mechanism to confine the quarks in a small region, we
will investigate the damping oscillation in next section more
detail. Therefore as the temperature is above $T_{C}$, the
confinement of the quarks is removed completely.

\begin{figure}
\includegraphics[scale=0.9]{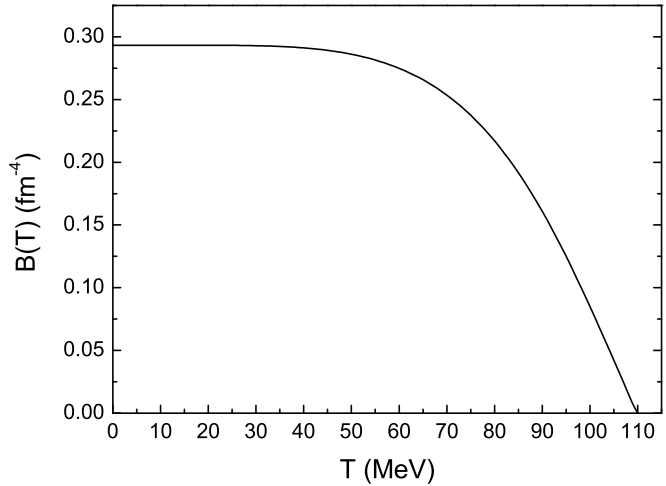}
\caption{\label{Fig:Fig6} The bag constant $B(T)$ as functions of
$T$.}
\end{figure}

\begin{figure}
\includegraphics[scale=0.9]{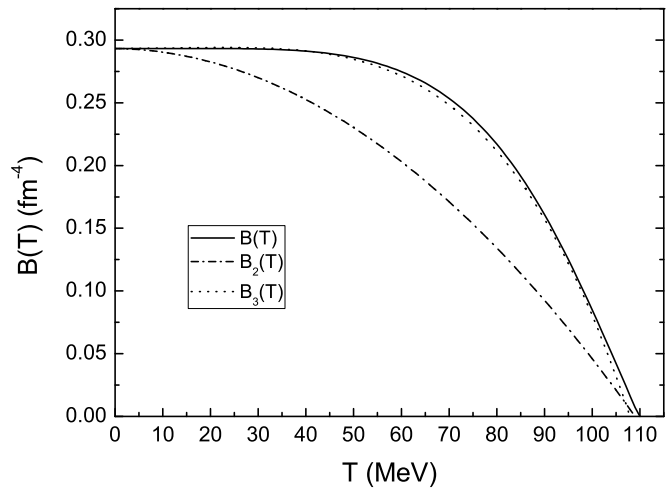}
\caption{\label{Fig:Fig7} The bag constant $B(T)$, $B_2(T)$ and
$B_3(T)$as functions of $T$. }
\end{figure}

Using the obtained temperature-dependent effective potentials, from
Eq.(\ref{bag}) we illustrate the temperature dependence of the bag
constant $B(T)$ in Fig.\ref{Fig:Fig6}, and it is shown that the bag
constant decreases continuously with increasing temperature. At the
critical temperature $T_{C}$, $B(T_{C})=0$. We can fit the
analytical formula of the bag constant as a function of $T$
\begin{eqnarray}
B_3(T)=B_0 \left[1-c1 \left(\frac{T}{T_{C}}\right)^2-c2
\left(\frac{T}{T_{C}}\right)^4 \right],
\end{eqnarray}
where $B_0$ is the bag constant at zero temperature, the parameters
are $c1=-0.113$ and $c2=1.113$ respectively. For comparison, we plot
$B(T)$, $B_2(T)$ of Eq.(\ref{e4}) and $B_3(T)$ in
Fig.\ref{Fig:Fig7}. $B_2(T)$ decreases smoothly with increasing
temperature, while $B(T)$ and $B_3(T)$ decrease dramatically with
increasing temperature when the temperature is around $85
\mathrm{MeV}$. From Eq.(\ref{e3}), we can give the parameters
$A_1\simeq -0.67$ and $A_2\simeq 1.55$.

\section{The soliton solutions and deconfinement phase transtions}

Soliton models such as those described in Section II can be modified
so as to allow for the effect of thermal background with temperature
$\beta$, just by replacing the relevant classical potential function
$U(\sigma)$ by an appropriately modified temperature-dependent
effective potential
$V(\sigma;\beta)$\cite{Holman:1992rv}\cite{Carter:2002te}.

For finite temperature soliton solutions will take the same form as
the one at zero temperature. But the equation of motion for the
$\sigma$ field should be replaced as
\begin{eqnarray}\label{equation31}
\frac{d^2
\sigma(r)}{dr^2}+\frac{2}{r}\frac{d\sigma(r)}{dr}-\frac{dV(\sigma;\beta)}{d\sigma}=Ng(u^2(r)-v^2(r)),
\end{eqnarray}
where the temperature-dependent effective potential
$V(\sigma;\beta)$ is defined in Eq.(\ref{potential0}). And the quark
functions should also satisfy the normalisation condition
\begin{eqnarray}
4\pi \int r^2 (u^2(r)+v^2(r))dr=1
\end{eqnarray}

The functions $\sigma(r)$, $u(r)$ and $v(r)$ also satisfy the
boundary conditions following from the requirement of finite energy:
\begin{eqnarray} \label{condition1}
v(0)&=& 0,   \frac{d\sigma(0)}{dr}=0, \label{condition1}\\
u(\infty)&=& 0.\label{condition2}
\end{eqnarray}
However, the situation is changed for $\sigma(r)$ as $r\rightarrow
\infty$. When $T\leq T_{C}$, in order to satisfy the requirement of
finite energy of the solition (or other topological defects), as
$r\rightarrow \infty$,
 $\sigma(r)$ should be equal to $\sigma_v$, where the potential
$V(\sigma)$ has an absolute minimum. For example, at zero
temperature, we take the asymptotic value (vacuum values) $\sigma
\rightarrow \sigma_v$ as $r\rightarrow \infty$, while for finite
temperature, $\sigma \rightarrow \sigma_v(\beta)$ as $r\rightarrow
\infty$, because $\sigma_v$ is a temperature-dependent function, and
of course, $V(\sigma_v(\beta);\beta)$ has an absolute minimum. For
$T> T_{C}$, the physical vacuum becomes unstable, and the stable
vacuum  is the perturbative vacuum which is the absolute minimum of
the effective potential. Therefore, to satisfy the requirement of
finite energy of the solition at $T> T_{C}$, we should take the
asymptotic value (vacuum values) $\sigma \rightarrow 0$ as
$r\rightarrow \infty$. Based on above analysis, one obtain the
following boundary condition for the function $\sigma(r)$ as
$r\rightarrow \infty$:
\begin{eqnarray}
\sigma &=& \sigma_v(\beta), \qquad \mathrm{for}\qquad T\leq T_{C}, \label{condition3}\\
\sigma &=& \sigma_0, \qquad \mathrm{for}\qquad
T_{C}<T.\label{condition4}
\end{eqnarray}

\begin{figure}
\includegraphics[scale=0.9]{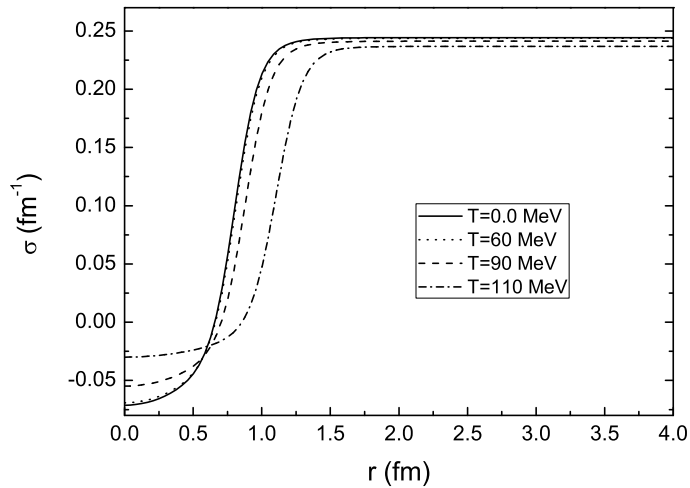}
\caption{\label{Fig:Fig4} The soliton solutions for different
temperatures when $T\leq T_{C}$. }
\end{figure}

\begin{figure}
\includegraphics[scale=0.9]{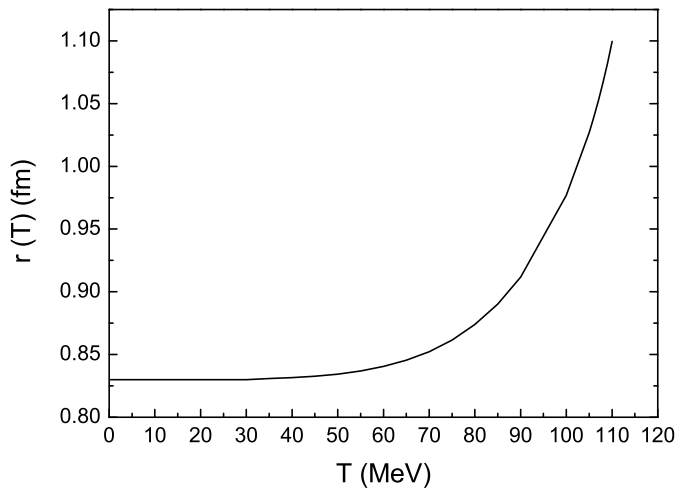}
\caption{\label{Fig:Fig8} Radius as functions of temperature when $T
\leq T_{C}$.}
\end{figure}

The set of coupled nonlinear differential equations
(\ref{equation1}), (\ref{equation2}) and (\ref{equation31}) can be
solved numerically with the boundary conditions
Eqs.(\ref{condition1}),(\ref{condition2}) and (\ref{condition3})
when the temperature is $T\leq T_{C}$. In Fig.\ref{Fig:Fig4}, we
plot the soliton solutions by taking the temperatures as $T=0
\mathrm{MeV}, 60 \mathrm{MeV}, 90 \mathrm{MeV}$ and $110
\mathrm{MeV}$. From Fig.\ref{Fig:Fig4}, we can see that with the
temperature increasing, $\sigma_v(\beta)$ is nearly constant near
$\sigma_v$, while $\sigma(0)$ is changed dramatically. Similarly to
the discussions of Ref.\cite{Zhang:2001ih}, where they predicted
that the radius of a stable strangelet increases as temperature
increases, we can give an exact variation of radius as functions of
temperature in Fig.\ref{Fig:Fig8}, which reveals that the radius of
the soliton bag does increase with increasing temperature.

\begin{figure}
\includegraphics[scale=0.9]{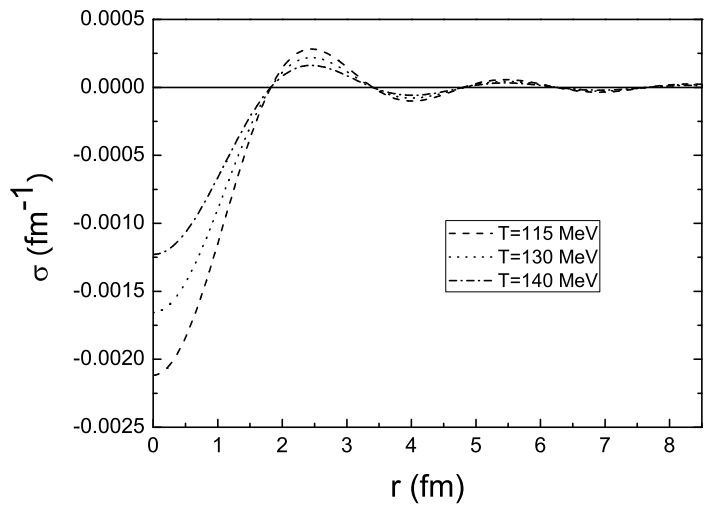}
\caption{\label{Fig:Fig5} The solutions for the set of coupled
nonlinear differential equations (\ref{equation1}),
(\ref{equation2}) and (\ref{equation31}) for different temperatures
when $T_{C}<T$.}
\end{figure}

\begin{figure}
\includegraphics[scale=0.9]{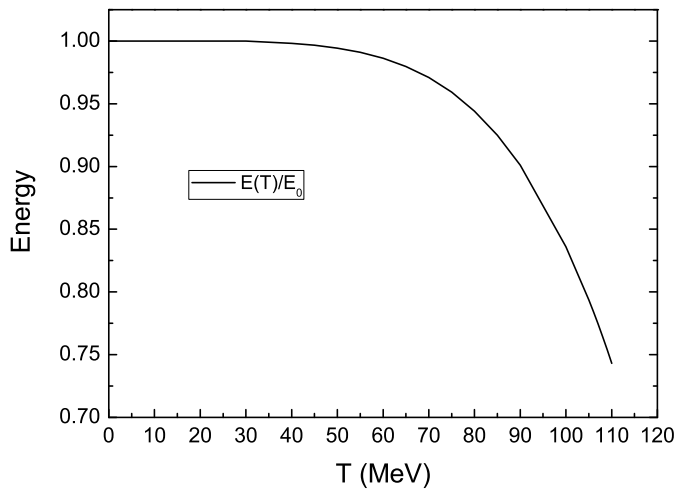}
\caption{\label{Fig:Fig9} The total energy of system (soliton mass
or energy)as functions of temperature when $0\leq T \leq T_C$,
where$E_0$ is the soltion energy at zero temperature. }
\end{figure}

Similarly, the set of coupled nonlinear differential equations
(\ref{equation1}), (\ref{equation2}) and (\ref{equation31}) can be
solved numerically with the boundary conditions
Eqs.(\ref{condition1}),(\ref{condition2}) and (\ref{condition4})
when the temperature is $T_{C}<T$. In Fig.\ref{Fig:Fig5}, we plot
the solutions by taking the temperatures as $T=115 \mathrm{MeV}, 130
\mathrm{MeV}$ and $140 \mathrm{MeV}$. When the temperature is higher
than the critical temperature $T_{C}$, the bag constant is zero, the
physical vacuum becomes unstable, and the perturbative vacuum state
is stable, then the shapes of solutions are very different from that
of $T\leq T_{C}$. With the increase of temperature, the soliton
solutions tend to disappear. From Fig.\ref{Fig:Fig5} we can see
that, unlike the conventional soliton solutions ploted in
Fig.\ref{Fig:Fig4}, here the solutions have the damping oscillation,
and we can not find soliton solution anymore.

For investigating the stability of solion at different temperatures,
it is necessary to calculate the total energy of the system at
various temperatures by using Eq.(\ref{energy}), accordingly the
relevant classical potential function $U(\sigma)$ should be replaced
by the temperature-dependent effective potential $V(\sigma; \beta)$.
In Fig.\ref{Fig:Fig9}, We plot the total energy of the system
(soliton mass or energy) as functions of temperature for $0\leq T
\leq T_C$, where$E_0$ is the soltion energy at zero temperature. It
can be seen that the total energy of system will decrease as the
temperature increases. When the temperature is around the critical
temperature $T_{C}$, the total energy of the system will
dramatically decrease to 74\% of $E_0$. Since there is no soltion
solution when $T>T_C$, the soltion mass (or energy) for $T>T_C$ does
not exist. However, from the Fig.\ref{Fig:Fig5} and
Eq.(\ref{energy}), we can find that the total energy of system is
almost equal to the quark field energy, this also implies that the
soltion bag is melted out and has disappeared when the temperature
is above the critical temperature $T_C$. Since there is no soliton
bag when the temperature is above the critical temperature $T_C$,
there is no mean-square charge radius for the proton for $T>T_C$.

As mentioned in Sect.III,  $B(T)$ is defined as the difference
between the vacua inside and outside the solion bag. For $T\leq
T_{C}$, the vacuum inside the soliton bag is the metastable vacuum,
the vacuum outside the soliton bag is the real physical vacuum. This
can be seen in Fig.\ref{Fig:Fig4}. As the temperature is above the
critical temperature $T_{C}$, the metastable vacuum becomes the
absolute one.  It can be seen from Fig.\ref{Fig:Fig5}, the behavior
of $\sigma(r)$ changes dramatically when $T>T_{C}$. $\sigma(r)$ is
very close to zero at any $r$. Therefore, the values of the
effective potential are always close to its stable minimum at
$\sigma=0$. The state at $\sigma \sim \sigma_v(\beta)$ can never be
realized in this case. This means that the bag constant $B=0$.

Based on above analysis, there is no soliton solution and the bag
constant $B(T)$ is zero for $T>T_{C}$, so there exists no more
mechanism in soliton bag model to confine the quarks, and the quark
is to be decofined when $T>T_C$.

\section{Summary and discussion}

In this paper, we have investigated the deconfinement phase
transition of the IQMDD model at finite temperature and obtained the
effective potential of the IQMDD model at different temperatures. We
also have gotten the function $B(T)$ and fitted the analytical
formula of the bag constant. It is shown that the two minima of the
potential at zero temperature will be equal at a certain temperature
$T_{C}\simeq 110 \mathrm{MeV}$. When the temperature $T>T_{C}$, the
original perturbative vacuum state becomes stable and the original
physical vacuum state becomes a metastable one. Because $\sigma(r)$
is very close to zero at any $r$ for $T>T_{C}$, the effective
potential takes always its value at the absolute minimum $V(\sigma
\sim 0)$. This gives $B(T)=0$ for $T>T_{C}$. As the temperature
approaches another higher temperature $T \simeq 150 \mathrm{MeV}$,
the effective potential has a unique minimum. Our results are
qualitatively similar to that obtained in conventional soltion bag
model at finite temperature\cite{Gao:1992zd, Li:1987wb,
Wang:1989ex}.

In IQMDD model, the confinement of the quark requires the existence
of the soliton solution, and the latter depends on the effective
potential at finite temperature. In order to investigate the
behavior of the soliton solution at finite temperature, we
numerically solve the set of coupled nonlinear differential
equations. Our results show that when $T\leq T_{C}$, there exist the
stable soliton solutions in IQMDD model, but when the temperature is
higher than the critical temperature $T_{C}\simeq 110 \mathrm{MeV}$,
there is only  damping oscillation solutions and no soliton-like
solution exists, and the quarks can not be confined by such
solutions anymore, then the confinement of quarks are removed and
the deconfined phase transition takes place. We also obtain that the
radius of the soliton bag does increase when the temperature
increases. At $T=T_C$ the soltion bag disappears.

Like the conventional soltion bag model, there are four adjustable
parameters in the IQMDD model and the numerical results are also
parameter-dependent. Another shortcoming of these bag models is lack
of chiral symmetry. This symmetry suggests the use of a linear sigma
model, which couples the quarks to pion fields as well as a scalar
field. Solitons in such a model have been studied by Birse and
Banerjee\cite{Birse:1983gm}\cite{Birse:1984js} and Kahana, Ripka and
Soni\cite{Kahana:1984dx}. It is of interest to extend their work to
finite temperature and discuss the soliton solutions at different
temperatures. Unlike the QMDTD model, where the bag constant B is
put in by hand through an ansatz, we have obtained the bag constant
B as a function of temperature, so our work can be used to study the
effect of $s$ quarks for strange quark matter. All these works are
in progress.

\begin{acknowledgments}
The authors wish to thank Jiarong Li and Song Shu for useful
discussions and correspondence. This work is supported in part by
the National Natural Science Foundation of People's Republic of
China.
\end{acknowledgments}

\end{document}